# Mutation Testing framework for Machine Learning

*Raju
Arizona State University, Tempe, US
Email: rsingh80@asu.edu
Phone: +91.997.211.8300
ORCID: 0000-0002-9522-1783

**ABSTRACT**
*This is an article or technical note which is intended to provides an insight journey of Machine Learning Systems (MLS) testing, its evolution, current paradigm and future work. Machine Learning Models, used in critical applications such as healthcare industry[1], Automobile [2], [3] and Air Traffic control, Share Trading etc., and failure of ML Model can lead to severe consequences in terms of loss of life or property. To remediate this, developers, scientists, and ML community around the world, must build a highly reliable test architecture for critical ML application. At the very foundation layer, any test model must satisfy the core testing attributes such as test properties and its components. This attribute comes from the software engineering [5], [6], but the same cannot be applied in as-is form to the ML testing and we will tell you "why".*

## 1. INTRODUCTION

In current context of software development and machine learning, it is inevitable, not to come across a machine learning scenario in day to day life. It spans across business critical applications such as share trading, insurance, banking, Medical applications such as - drug manufacturing, identification of disease, medical imaging, Safety critical applications such as autonomous driving, robotics. Applications of ML in several critical sectors makes ML testing a reliable way to ensure quality and minimize failure scenarios. An adoption of testing framework from traditional software engineering and testing with addition of key ML quality attributes make more sense. Testing framework which covers performance (for critical real time systems), security (for business applications, health care applications), safety (for system systems) increases the trustworthiness of it.

To better understand the testing challenges for ML systems, we need to deep dive as how ML systems are different from traditional software system. Traditional software are more deterministic in nature, lacks dynamicity in-terms of varied inputs. On the other hand, ML systems are dynamic, non-deterministic and expected to learn from data (labels) and predict the output accordingly. Example, a rover has to determine the path on a rocky terrain based on the imaging data that it gathers from the surrounding, the forest fire alert systems has to generate a prediction based on the environmental data such as air humidity, wind flow [direction, temperature, climatic conditions. The model tends to evolve and learn from historical data.

*Oracle Problem [7]*: Machine learning models are difficult to test, and it is because the it is designed to solve problem based on learning from past experience (label data, supervised learning), or without past experience (unsupervised learning) or through re-enforcement. Attempts has been made to draw parallel between the ML testing approach with Software Testing. By understanding the process of Software development, we should be able to break down the software stack into components (unifiable units), and build test cases around it. In other approaches, we have test driven development to setup in testing framework. This approach might now work well with ML models. It is because, machine learning models are mostly monolith, and components may not reflect the true nature of the ML model as a whole. Again, breaking ML model into unifiable components or developing with TDD is a cumbersome task.

In order to understand and design a test framework for ML system, we need to understand the behaviour [4] of the model, and how the model interacts with the surrounding. Studying behaviour of the ML model, gives limited insight into the model, not it is reasonably well to start with.

## 2. Background

[Def: Machine Learning: Machine learning is a field of study that gives the ability to computers to learn without being explicitly programmed. A computer program is said to learn from experience E with respect to some task T and some performance measure P, if its performance on T, as measured by P, improves with experience E. [12], [13], [23].]

Machine learning is a phased approach. The first phase is the learning phase. In this phase, data is gathered and bucketized as training and test data sets. Training data set is identified by attributes and label. The outcome of this phase is a model that is drawing the relationship between the attributes and the label. The subsequent phase deals with applying the model to different dataset (test data). There are several algorithm to accomplish this, such as classification algorithm, ranking algorithm etc.

Terms used in machine learning domain:

*Dataset*: An ingredient for machine learning model, consists of sets of instances for building or evaluating the model. It is further categorized as:

*Training Data*: This data is obtained from the sources (sensors, data collection devices etc., aggregated and cleaned up to exclude bias and noise) and is used for training purpose a machine learning mode. This model is basically an organic algorithms which learns from the training data and performs a particular tasks.

*Validation data*: This data is from the training data, used to tune the hyperparameters of learning algorithm.

*Test data*: This data is the part of training data, for which machine learning model has not been trained yet. Based on the performance of the machine learning model, and its behavioural with the test data, we can attribute the machine learning model maturity.

Sub Definitions

*Instance* is an information record about the object. *Feature* is a measurable property. Error are also an important aspect of the machine learning, and it is this property on which the model behaviour depends. *Test error* is mainly focused on deviation measure between the obtained value and the expected value.

Let us classify the Machine Learning

*Supervised Learning[19]*: The goal is to predict the value of an outcome measure based on the a number of input measures. It is commonly referred as regression [16] problem as its the outcome measurement is quantitative.

*Unsupervised Learning[18]*: The goal is to describe the association and patterns among a set of input measures. We only observe the feature and have no measurement about the outcome.

*Reinforcement Learning[20]*: This is a different approach, critical and has immense potential for prediction models. The agent (learning system) can observe an environment, selectively perform actions, and get rewards (or penalties). The key here is, it must learn it my itself, and accordingly responds in actions for the good. This approach is referred as policy, to get most reward over a period of time. In nutshell, a policy defines what actions the agent should take when subjected to a condition.

### 3. ML Testing Scenarios

Fault [8], [9] and Failures: Prior discussion involves classification of the faults and failure scenarios of the ML systems. Since most of the ML systems deals with uncertainty components, faults and failure are possible in ML models. This can be handled by creating counter measures to prevent failure scenarios, however, we can have unavoidable scenarios.

Definitions in the IEEE Standard Glossary (IEEE 1990):

Def [ Fault ] [24] An incorrect step, process, or data definition in a computer program.

Def [ Failure ] [24] The inability of a system or component to perform its required functions within specified performance requirements

Def [ Data Sensitive Fault ] [24] A fault that causes a failure in response to some particular pattern of data.

Def [ Program Sensitive Fault ] [24] A fault that causes a failure when some particular sequence of program steps in executed.

The core of the testing system is to find the deviation of ML models from the expected outcome.

Oracle: Oracle tests are basically intended towards the Behaviour test. This is a challenging aspect as the behaviour of ML systems are unpredictable, and this unpredictability makes sure sense to build oracle tests. In MLS context, Metamorphic Oracles have gain ground as a feasible approach to infer oracle information from data. Metamorphic oracles insights metamorphic relations between input values, i.e. if a metamorphic relation exists between the inputs, the corresponding MLS outputs must satisfy a pre-existing relation (ideally equality or equivalence relation). Input data and its dimension poses a greater instability towards the ML testing realm. So, it is vital to choose adequate test data in order to cover impactful dimensions.

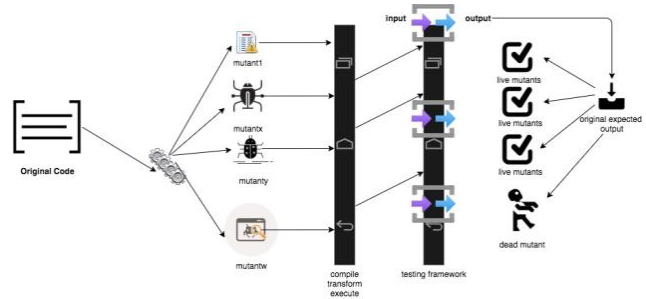

### 4. ML Testing

Software testing techniques can be used in ML testing domain such as unit, integration and system testing. In addition to this, in order to address the dynamicity of the ML systems, additional recommendation has been made for ML testing domain which includes – input, model, integration testing.

*Input testing*: This tests are concentrated on the input data which is used to train the ML model. The core reason of using input test is to minimise the faults risk. It can be either offline testing or online testing. During the offline testing it detects faults by alerting the bias in the training data. While in online testing, where ML models is expecting to predict for unlabelled data, this testing helps on input validation.

*Model testing*: Model testing tests the function aspects of the system under test (SUT) (ML model in isolation, without taking any other component into account). It tries to find the faults in the model architecture, training process etc. It uses accuracy (for classifier) or mean squared error (for regressions [16]). It is sometimes considered as unit testing.

*Integration testing*: Integration testing considers the integration aspect of ML modes, hardware systems, software systems and its interactions.

*System testing*: System testing is a holistic test to evaluation the systems measures under a given requirement.

*Black-box and White-box testing:* Black-box testing[21] screens the internal structure of the design, code and its implementation, of ML systems, without having access to the core while white-box testing is crucial as it knows the internal structure of the code, design, implementation and behaviour of the ML systems. This way it makes more sense to use white-box testing in ML model. In contemporary software, source code is the main source of faults or defects. Mutation testing injects modified program code to introduce defects or faults, and this enables the qualitative measurement of test data by detecting manual changes. With the knowledge on such mutation testing framework, we can suggest a DL based mutation testing framework with two stage process.

Source level mutation: DL systems is depends on the training program and training data. Training process is defined as the articulation of training program on training data. The master source code of training program and master record of the training data is mutated over a period of time during the testing and the deviation of the Model is recorded. The new evolved model, result from the mutation exercise, is set to run through the training set in order to determine the quality of the test data. The mutation operator can be categorized as: data mutation operator and program mutation operator..

*Data Mutation operators*: As the name suggest, DL model depends heavily on training data. We know that DL model robustness depends on the underlying data quality. Error introduces at any stage of data collection, data aggregation, data cleaning and skew the DL model as the data contain noise.

*Program Mutation Operators*: Training programs in DL systems are coded using high level languages, and uses problem specific programming framework. Injection faults in the program would causes unexpected behaviour in the DL systems.

This requires us to carefully craft mutants operations to inject faults into the training program. The kind of fault we can think of now is like, addition and removal of layers from DL models, pass on skewed weights and activation function while training process.

*Model Mutation Testing for DL systems*

Most of the mutation testing framework which works efficiently in traditional software systems doesn't hold ground with the DL models. The problem is, most of the mutation testing from traditional system is written on the source code, or its low level representation such as byte-code. However, Model mutation testing can be a better approach and we will show how it is it.

In *source level mutation testing*, the algorithm injects modifications in the training data and training program. While, in *model level mutation testing*, the algorithm updates the DL model obtained from the training program. As the expectation remains intact for both approaches, that is to evaluate effectiveness and weakness of the test data set, model level mutation testing's leads the way forward by directly mutating the DL model.

*ML System attributes*

Security: ML systems are as vulnerable as any other software systems, along with few inherited vulnerability, given the model footprint. Security is the reciprocates to the robustness.

Efficiency: ML system efficiency reciprocates to accuracy of its prediction.

Fairness: ML systems suffers from statistical problems such as bias, deviations, skewness.

## 5. ML Testing Framework

*Behaviour framework*: ML system might behave different given similar data. The main challenge is to identify the extreme boundaries for a given input space. This is similar to boundary-value analysis.

Test Adequacy criteria: Any test suite woven around an ML system, should satisfy the quality attributes. As the classical approaches (based on the source code control flow [10]) is not relevant to the ML systems, researchers are trying to find out new domain in order to satisfy the test adequacy.

*Mutation*: In contemporary software testing domain, mutation testing is gaining grounds. It is become an efficient tool to find the faults into the ML systems, by injecting mutants. DeepMutation fundamentally works at the model level, iterates through varying mutation within the boundary space.

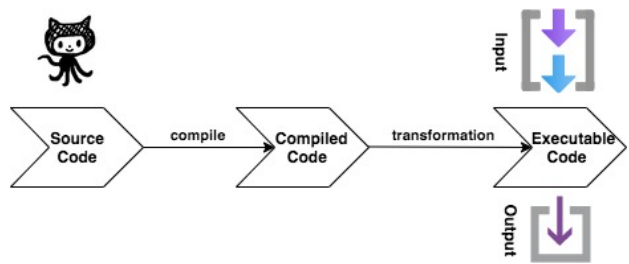

*DeepMutation*: Deep neural networks has gain ground in several critical applications such as healthcare, autonomous vehicle, robotics. Any DNN systems can either be a FNN or RNN systems. A FNN system process the input information at each layer and forwards its to the next layer. This process continues until the decision is reached. This way, the FNN model preserve the local properties of each layer. While the RNN extends the LSTM or memory cells and partially propagates the information backward to secure temporal information of sequential inputs. This way, the decision at any stage not only depends on the given input, but also on the current state. This makes RNN reliable for handling sequential data, example, NLP. The spirit of mutation in DNN is similar to that of traditional software. The main idea behind DeepMutation is to introduce adequate number of mutants or operators. The mutants must satisfy the quality attributes for the testing framework, such as input (test) data analysis.

Traditional software is built upon decision logic. This logic is implemented in form of program codes. While the DL models and systems are guided by of the underlying *Deep Neural Network (DNN)* structures and its weight.

The weight of DL system is generally obtained from executing training program on a given training data, and

DNN structure is defined as the code of the training program. These are two potential reason, a deviation in which cause behavioural issue in the DL systems. The mutation operator can be inflicted in either training data set of training program or both. Once the mutant operator are injected, training program is executed on training data to generate mutated DL models.

*DeepMutation Testing Framework*

DNN uses high level languages such as python, r, however DNN is represented as hierarchical data structure. We are going to shortly lay down on to discuss on the mutation testing framework for DL systems. The first steps is to design *source level mutation testing operators*. These operators can modifies the training data and training program. The basic idea behind this is to improve the data quality evaluation. The fault might be inject manually or naturally occur in the training data or in training program. This framework must address the mutated DL models efficiently and address issues such as computation resource requirements, security vulnerability issues. Given this, we must work backward to generate efficient mutant operators. Before we deliver further, we need to elaborate model-level testing.

*Model level testing:* A model is used to represent the desired behaviour of the system under test (SUT), or to represent the testing strategies and a test environment. A model representation of SUT is at abstraction or partial behaviour. We can derive only functional test cases from SUT. The idea here is to come up to the conclusion that how many model level mutation operators would result in the efficient generation of a set of mutations without inducing model level problem.

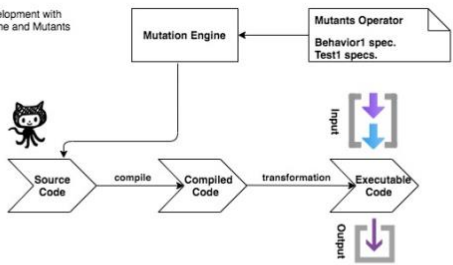

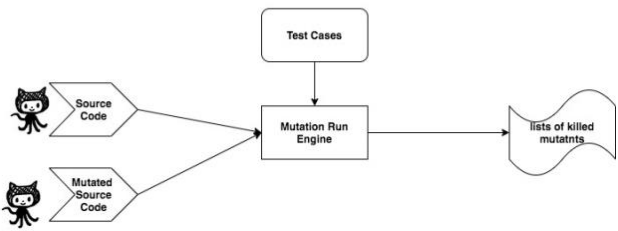

## 6. Prediction Mutation Testing

Prediction Mutation Testing topic tends to attract two short of discussion – Mutation Testing approach in Machine learning, and – Machine learning approach in mutation testing. We will talk a little about the later part, and then resume the discussion on the primary topic which is related to explore the possibility of testing a machine learning models such as DNN using mutation testing, its framework, challenges, pros and cons, future works etc.

When we discussion the approach of machine learning and its impact on mutation testing, we consider that here mutation testing can be again applied to a software system or machine learning systems. However, applying mutation testing in software system is inherently different from apply mutation testing in machine learning, and its because of the behavioral changes that software systems and machine learning systems exhibits when subjected to mutation testing.

Testing, in general is a powerful and unified (yet distributed) way to evaluate the quality of underlying systems, be it traditional software systems or machine learning systems (DNNs). In this approach, we tend to generate a large number of mutants and execute against a test suite to check the ratio of killed mutants. This makes mutation testing computationally expensive. So, it's worthwhile to invest some time to learn about predicting behavior of mutation testing. It is important to include here that, this approach is based on the classification model. This model predicts whether mutants are killed or survived during the testing without executing it. However, unlike several predicted problem, this approach also suffers from accuracy loss (which we can ignore as it is minimalistic).

In general, mutants are a set of program variants (or training program in machine learning). A set of transformation rules generates mutants from the original program. These mutants are called as mutation operator, that seed logic and syntactic changes into the program one at a time.

*Killed mutants*: A mutant, killed by a test-suite, if at-least of the test from the source has a varied execution behavior on the mutants and original program. Such mutants are called *killed mutants*. Elsewise, the mutant is known to have survived. The ratio of killed mutants to all mutants (non-equivalents) is referred as mutation score. It is usually used to evaluate the test suite's effectiveness.

Other areas, where use of mutation testing is prevalent are simulation testing, localizing faults, model transformation and guided test generation.

As mentioned earlier, mutation testing is an extremely expensive approach. It requires generating and executing each mutant are the test suite. Both of these activities – generation of mutations and execution of mutants are expensive operations on hardware of scale. In recent time however, we have seen phenomenal progress to bring down the operational cost for mutant generation, however, executions remain expensive in spit of several refinement techniques such as selective mutation testing, weak mutation testing, high-order mutation testing, optimized-mutation testing etc.

Due to the problem faced for the expense vs effectiveness, predictive mutation testing is gaining ground. Mutation testing enables machine learning to build a predictive model by means collecting a series of features These features can be test-suite coverage or mutation operators on already executed mutants of earlier versions of the project or even other projects.

Earlier versions of same projects are commonly referred us as cross-version prediction. Cross project predictions are referred to other projects under test.

*Tradeoff: Efficiency vs Effectiveness*

Any prediction model inherently suffers from accuracy problem. However, several experimental approaches in predication mutation testing domain has shown positive signs as it improves efficiency and accuracy of mutation testing. This is a clear indication of how prediction mutation testing stands out of traditional mutation testing. It will be worthwhile to look into the class probability distribution provided by the classifier, with which developer may choose the mutant with proper probability distribution to get better prediction result. It is a considerable improvement over traditional mutation testing, as its light weight, in-expensive comparatively, with relatively high accuracy. In this article, it is assumed that mutation testing, refers to prediction mutation testing.

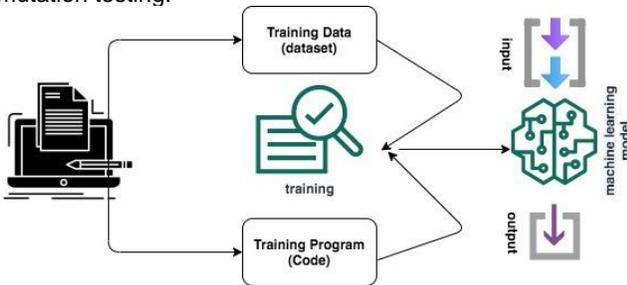

### 7. Model Mutation Testing

Models are common in software testing. It is used to select test suites. Applications of mutation testing at a model level can contribute to reliable and early assessment of the quality of the test suits. This can also help in defining a test suite which has high fault detection rates. One of the issues which we observed while using mutation testing[22] at the early development stage, is related to its reliability and quantifying it. There are several attempts and general-purpose availability of model based mutation testing, which depends upon comparing results from different test scenarios.

### 8. Challenges

Contemporary coding for functional requirements is different from programming a DL model. The basic different is, in contemporary programming, we break down the monolith requirements into small chunks of programmable units, each unit is programmed separately and satisfy the software quality attributes such as correctness, fairness, security etc and then it can be combines together with other modules to form the holistic program. In this approach, each unit or module has its own logic, an aggregation of which comply with the integrity attributes of the whole program.

While in DL systems, which fundamentally a data driven model, the logic that we can to drive at the highest may of abstraction might not be same when we try to modularize it. What is it so? Because the logic is guided by the weight and activation functions. Moreover, DL systems are behaviour driven systems which is built by executing training program on training data. Here, underlying logic is guided by the training data and not the requirement (as in traditional software).

### 9. Hypothesis

Let us make some assumption about the samples, adversarial samples and normal samples. In testing, adversarial samples are those samples which are vulnerable to any changes and shows far more deviation in behaviour with respect to usual samples. Consider a scenarios, the original DNN models has undergone a set of transformation rules to yield mutated DNN models. These mutated DNN models usually tends to label an adversarial data with a different label (label generated by original mutated DNN). We would assume this state, and try to measure following crucial factors such as model uncertainty estimate, density estimate, model sensitivity to the input changes.

Even before we can create a procedure for our hypothesis, we need an efficient way to generate the mutants. The fundamental approach is to generate or seed several program level mutation (mutants). This would require program under consideration to go through set of mutation operators by apply set of transformation rules. To the core of which lies, the process to define mutants operators. As if is known that traditional software systems are logic oriented, structured, while, DNN models are behaviour and model oriented, mutation operators application in the former scenario (Traditional Software Systems) is not application to the latter (DNN Systems). There are quite a few techniques which works independently and using mutation testing in order to establish the testing framework.

### 10. Approach

*Initialize*

One of the initial approach would be a way to foundation steps for measuring label change rate, also known as LCR, for the adversarial samples and normal samples. This is measurable when we inject these samples into a set of already mutated DNN models.

Table 1: Model Mutation Operator

| Mutation Operator | Level | Description |
|---|---|---|
| Gaussian Fuzzing (GF) | Weight | Fuzz weight by Gaussian Distribution |
| Weight Shuffling (WS) | Neuron | Shuffle selected weights |
| Neuron Switch (NS) | Neuron | Switch two neurons within a layer |
| Neuron Activation Inverse (NAI) | Neuron | Change the activation status of a neuron |

*x: input sample* (adversarial sample or normal sample).

*f: DNN model* (post mutation operators are applied).

Now, we go through the model mutation operator as provided in the Table 1 (sequence wise) and select the mutation models. Quite few times, the output mutated model is of moderate to low quality (assuming high precision and confidence as measure of high quality mutated models). This means that the accuracy and effectiveness on the training data works well, however on the test data it significantly deprecates. We let go or ignore these low quality mutated models. Only mutated models with high accuracy is considered. We can adopt the scale based on our experience and historical data obtained from mutated models. Ideally, any model with more that 90% accuracy of the original model is part of the set. This is to make sure that we meet the decision boundary [15] conditions and it is not impacted or much. Upon segregating the mutated models, we further obtain a label of the input sample on each mutated model.

*Build a Model*

In this stage, we follow the hypothesis to create a model. This model validates (on certain criteria) the observation. If we recount, earlier we mentioned that adversarial samples are generated in such a way that it tends to minimize the mutated behaviour on normal samples, while, it is being able to jump the decision boundary [15]. There are different ways of mutation to achieve this behaviour. As per the hypothesis, the effective adversarial samples are closer to the decision boundary. This minimizes the restricted modification in the model. With this, adversarial samples would be considered as a case of cross the decision boundary, unlike randomly selected mutated model. This implies, if we inject mutated adversarial sample into the mutated model, the outcome of the label tends to change it from its original label.

*Algorithm Design*

Experiments and test results shows that LCR can be a distinguisher between adversarial samples and normal samples. We can discuss on the algorithm which can be designed to detect samples at runtime based on LCR measures of a provided samples. This algorithm would delete the LCR and keep on generating more effective and accurate mutated models. For this to happen, we must define a satisfiable stopping condition on the mutation model generation algorithm. Prediction algorithm can helps us get a set of mutated models with higher accuracy before-hand.

## 11. RELATED WORK

Mutation testing in traditional software predates any similar testing framework in Machine Learning (DNN specifically). There, mutation testing a proven tool and has higher accuracy. Mutant operators is a well-researched area and it has its implementation in several high level programming languages for traditional software. With time, and increasing complexity of traditional software, the mutation testing framework has been well extended. Coming to the use mutation testing in machine learning area, it is being researched and several researchers have established milestone for most of the known DNN modes. One such approach is DeepMutation.

## 12. CONCLUSION AND FUTURE WOR

In this work, we propose the machine learning mutation testing framework, its usefulness and approach to detect adversarial samples for DNN at runtime. We laid down the details of source level mutation techniques on datasets (training and test) and training (or test) programs. This required us to further the details the process and techniques involved in designing source level mutation operators, and feed faults into the DNN models during its development and testing process. This accompanied the details of model level mutation technique. Model level mutation technique differs from the source level mutation technique in the approach that it adopts to inject the faults. Model level mutation techniques directly feeds the faults into DNN system. It is also noteworthy about how to measure the quality of these mutation models.

We also briefly touched upon how to predict the mutants operators even before we can analyses it by executing. This is primarily done as mutants generation is an computationally expensive approach. In the end, we proposed a hypothesis and approach to build the problem set, analyse it and proceed under certain assumption to mitigate it.

## 13. Compliance with Ethical Standards

**Conflict of interest** The authors of this article state that there are no conflicts of interest.